# A study of local and non-local spatial densities in quantum field theory

R.E. Wagner[1], M.R. Ware[1], E.V. Stefanovich[2], Q. Su[1,3] and R. Grobe[1,4]

(1) Intense Laser Physics Theory Unit and Department of Physics,
Illinois State University, Normal, IL 61790-4560 USA
(2) 2255 Showers Drive, Unit 153, Mountain View, CA 94040 USA
(3) Beijing National Laboratory for Condensed Matter Physics, Institute of Physics, Chinese
Academy of Sciences, Beijing 100190, China
(4) Max-Planck-Institut für Kernphysik, Saupfercheckweg 1, D-69117 Heidelberg, Germany

We use a one-dimensional model system to compare the predictions of two different "yardsticks" to compute the position of a particle from its quantum field theoretical state. Based on the first yardstick (defined by the Newton-Wigner position operator), the spatial density can be arbitrarily narrow and its time-evolution is superluminal for short time intervals. Furthermore, two spatially distant particles might be able to interact with each other outside the light cone, which is manifested by an asymmetric spreading of the spatial density. The second yardstick (defined by the quantum field operator) does not permit localized states and the time evolution is subluminal.

## 1. Introduction

A quantum field theoretical system is called local if the field operator $\hat{\int}(z)$ in the interaction energy density with argument $z$ is coupled to other fields or itself at precisely the same variable. [1] As a result any two physical objects that are far apart and described by the field operators should not be able to interact instantly, reflecting the absence of any action at a distance. While quantum mechanically entangled particles can violate this principle [2], it is presently believed that this phenomenon cannot be used to transport any information or particles with velocities that exceed the speed of light. Equivalently, two measurements with a space-like separation should be independent of each other and the corresponding observables should commute. So far all experiments are consistent with this principle and any action at a distance has not been observed.

In this note we would like to point out that the above discussion relies on a particular interpretation of the argument $z$ and also on an assumption about the nature of a spatially localized state. This state should be defined as an eigenstate of the position-operator. However, even in an interaction-free quantum field theory this state is in general not necessarily given by the action of the field operator $\hat{\int}(z)$ or its adjoint on the vacuum state, $\hat{\int}^{\dagger}(z)|\text{vac}\rangle$. The requirement that position eigenstates with different eigenvalues z should be orthogonal to each other is violated for these particular states, in other words, $\langle\text{vac}|\hat{\int}(z_2)\,\hat{\int}^{\dagger}(z_1)|\text{vac}\rangle$ does not necessarily vanish for $z_1 \neq z_2$. This unfortunate state of affairs was already recognized early on [3] when it was recommended that possibly only products of field operators averaged over finite regions in space might have a physically observable meaning. This restriction was associated with a limitation of the continuous-field description that provides an adequate description of the world only for large spatial intervals. One could also argue that the argument $z$ of the field operator is merely an abstract integration parameter that is not necessarily related to the physical position.

Alternatively, a different concept for a position operator has been proposed [4] that permits localized and therefore mutually orthogonal states. This so-called Newton-Wigner operator has led to a long debate concerning which of the two proposals is better suited to describe the physical measurement of a particle's position. A clarification of this open question is even more desirable now as there has been a significant amount of work devoted to the analysis of the quantum mechanical dynamics [5] in the relativistic regime with full spatial resolution. These studies have included the spatial details of the ionization of atoms and ions by very strong external fields, the generation of higher harmonics and the supercritical field induced breakdown of the vacuum with the generation of electron-positron pairs. As some of the predictions become more and more accurate, it is important to understand how to calculate the particles' position accurately. As experiments are also entering the relativistic regime, it is essential that the abstract debate about the relativistic localization problem is shifted to a more quantitative analysis with the ultimate goal to develop concrete predictions that permit experiments to discriminate between both concepts.

In this work we will restrict the spatial dimension to one. This approximation can be quite serious, if phenomena are investigated that are intrinsically three dimensional in nature, such as the motion of a charge in an electromagnetic field. However, in many cases this restriction is not so serious and can permit a first qualitative insight and valuable intuition in complicated dynamical processes whose description in all dimensions is mathematically and computationally too difficult. In the early sixties a ground breaking work by Eberly [6] showed that even the concept of partial wave decomposition and the optical theorem have their direct counterpart in two and even one spatial dimension.

In some cases, due to the symmetries of the physical situation there is sometimes a dominant spatial direction permitting us to neglect the other two spatial dimensions as a good approximation. For example, more than fourty five different research groups [7] have modeled the ionization dynamics of atoms in strong laser fields using this dimensional restriction. These contributions led to several suggestions for the mechanisms of above-threshold ionization, higher-harmonics generation, stabilization and various multi-electron ionization paths.

In this work we use quantum field theory in one spatial direction, and so far none of the qualitative conclusions about the time-evolution of spatial densities, their localization or superlumimal behavior depends on the spatial dimension. For a comprehensive review on (1+1) dimensional quantum fields theories, see e.g. [8]. Obviously, due to the larger phase space, force laws for one-dimensional systems usually have different scaling properties with respect to the inter-particle spacing, but nevertheless fundamental aspects of the particle dynamics can be obtained with these toy models. For example, the role of particle dressing, locality, correlation and other properties for the time evolution of interacting physical particles can be examined with the hope of generalization of these findings to three-dimensional world.

It is our goal to contribute to this debate about the position operator by illustrating the different consequences of these two position yardsticks for a concrete and numerically tractable model system. In order to examine the properties of both position operators with regard to locality and action at a distance, we study in this note the one-dimensional (relativistically invariant) $\hat{\varphi}^4$-system.  We will show that in the interaction-free limit an initially localized particle (meaning a state of finite spatial support) can spread instantly to all regions in space according to the second yardstick.  This superluminal propagation raises the possibility of permitting two space-like separated particles to interact instantly with each other, which would violate the usual interpretation of the principle of causality.  In initiating a discussion of this non-trivial issue, we derive how these yardsticks are transformed for a velocity shifted coordinate frame.  We finish this work with a rather extended outlook into future work.

## 2. The model system

In order to have a concrete example to make numerical predictions for the two position yardsticks, we choose neutral scalar bosons of (bare) mass m in one spatial dimension.  Throughout this article, we employ atomic units where the speed of light c=137 a.u., the electron's mass and charge m=e=1 a.u. and $\hbar$=1 a.u.  In order to be able to study the interaction between particles as well, we will include a $\hat{\varphi}^4$-interaction with coupling strength $\lambda$ in Section 5.  The relativistically invariant Hamiltonian density (after renormalization) is given by [9-11]

$$\hat{H}(z) = \tfrac{1}{2}\, c^2\, \hat{\Pi}(z)^2 \; + \tfrac{1}{2}\big(\partial_z \hat{\varphi}(z)\big)^2 + \tfrac{1}{2}(mc)^2\, \hat{\varphi}(z)^2 + \lambda : \hat{\varphi}(z)^4 : \qquad (2.1)$$

Here we denote with the colons the normal ordered products with respect to the momentum operators $\hat{a}$, such that $:\hat{a}(p_1)\hat{a}^\dagger(p_2): = \hat{a}^\dagger(p_2)\,\hat{a}(p_1)$.  The real quantum field operator $\hat{\varphi}$ and its canonical momentum $\hat{\Pi}$ have to satisfy the required equal-time commutator relationship $[\hat{\varphi}(z_1), \hat{\Pi}(z_2)]_- = i\delta(z_1-z_2)$, where $z$ in italics denotes the (one-dimensional) argument which has the units of length.  In terms of the usual momentum annihilation operators $\hat{a}(p)$ they can be expanded as

$$\hat{\varphi}(z) \equiv (4\pi)^{-1/2} c \int dp\, \omega(p)^{-1/2}\, [\hat{a}(p)\, \exp(ipz) + \hat{a}^\dagger(p)\, \exp(-ipz)] \qquad (2.2a)$$

$$\hat{\Pi}(z) \equiv - i\, c^{-1}\, (4\pi)^{-1/2} \int dp\, \omega(p)^{1/2}\, [\hat{a}(p)\, \exp(ipz) - \hat{a}^\dagger(p)\, \exp(-ipz)] \qquad (2.2b)$$

where $[\hat{a}(p_1), \hat{a}(p_2)^\dagger]_- = \delta(p_1-p_2)$ and the bare energy $\omega(p) \equiv \sqrt{m^2 c^4 + c^2 p^2}$.

When we integrate the energy density operator $\hat{H}(z)$ over the variable $z$, we obtain the quantum field theoretical Hamiltonian $\hat{H}$. For the discussion below, the Fourier transform of the momentum operator $\hat{a}(p)$, defined as $\hat{a}(z) \equiv (2\pi)^{-1/2} \int dp\, \hat{a}(p) \exp(ipz)$, is important. Note here and from now on the argument z is purposely not typed in italics. The necessity for this seemingly irrelevant distinction between the arguments of $\hat{\varphi}(z)$ and $\hat{a}(z)$ will be clear below. The Hamiltonian $\hat{H} = \hat{H}_0 + \hat{V}$ can then be expressed in terms of either the fields $\hat{\varphi}(z)$ and $\hat{\Pi}(z)$, or equivalently in terms of $\hat{a}(z)$ and $\hat{a}^\dagger(z)$ as

$$\hat{H}_0 \equiv \int dz\, \{\tfrac{1}{2} c^2 \hat{\Pi}(z)^2 + \tfrac{1}{2}(\partial_z \hat{\varphi}(z))^2 + \tfrac{1}{2}(mc)^2 \hat{\varphi}(z)^2\} = \iint dz_1 dz_2\, V_1(z_1, z_2)\, \hat{a}^\dagger(z_1)\hat{a}(z_2) \qquad (2.3a)$$

$$\hat{V} \equiv \int dz\, \lambda : \hat{\varphi}(z)^4 : = \iiiint dz_1 dz_2 dz_3 dz_4\, V_2(z_1, z_2, z_3, z_4) \times$$
$$\times \big[ \hat{a}^\dagger(z_1)\hat{a}^\dagger(z_2)\hat{a}^\dagger(z_3)\hat{a}^\dagger(z_4) + 4\, \hat{a}^\dagger(z_1)\hat{a}^\dagger(z_2)\hat{a}^\dagger(z_3)\hat{a}(z_4) + 6\hat{a}^\dagger(z_1)\hat{a}^\dagger(z_2)\hat{a}(z_3)\hat{a}(z_4) +$$
$$+ 4\, \hat{a}^\dagger(z_1)\hat{a}(z_2)\hat{a}(z_3)\hat{a}(z_4) + \hat{a}(z_1)\hat{a}(z_2)\hat{a}(z_3)\hat{a}(z_4) \big] \qquad (2.3b)$$

The couplings between different variables for a single particle $V_1$ and between several particles $V_2$ are given by

$$V_1(z_1, z_2) \equiv 2\, c^2 \int dz\, I_{1/2}(z-z_1)\, I_{1/2}(z-z_2) = (2\pi)^{-1} \int dp\, \omega(p) \exp[ip(z_1-z_2)] \qquad (2.4a)$$

$$V_2(z_1, z_2, z_3, z_4) \equiv \lambda \int dz\, L_{1/2}(z-z_1)\, L_{1/2}(z-z_2)\, L_{1/2}(z-z_3)\, L_{1/2}(z-z_4) \qquad (2.4b)$$

Here the two integration kernels are defined as

$$L_{1/2}(z) \equiv c\, 2^{-3/2}\, \pi^{-1} \int dp\, \omega(p)^{-1/2} \exp(ipz) \qquad (2.5a)$$

$$I_{1/2}(z) \equiv c^{-1}\, 2^{-3/2}\, \pi^{-1} \int dp\, \omega(p)^{1/2} \exp(ipz) \qquad (2.5b)$$

While the first function $L_{1/2}(z)$ is real and positive and can be expressed in terms of a modified Bessel function, the second function $I_{1/2}(z)$ is complex and formally infinite. Note that the two functions also fulfill the useful orthogonality relationship $2\int dz\, L_{1/2}(z-a)\, I_{1/2}(z-b) = \delta(a-b)$.

We finish this section by comparing the equation of motion for $\hat{\varphi}(z,t)$ and $\hat{a}(z,t)$. While the time-evolution for both operators is given by the Heisenberg equation $i\,\partial\hat{A}(z,t)/\partial t=[\hat{A}(z,t),\hat{H}]_-$, we point out that for $\hat{A}(z,t)=\hat{a}(z,t)$ it reduces in the $\lambda\rightarrow 0$ limit to the relativistic Schrödinger-like equation [12,13], $i\,\partial\hat{a}(z,t)/\partial t=\sqrt{[m^2c^4-c^2(\partial/\partial z)^2]}\,\hat{a}(z,t)$, with the non-local square-root operator. This shows the direct relationship between the Klein-Gordon equation and the relativistic Schrödinger equation. The field $\hat{\varphi}(z,t)$ remains real under its time evolution and satisfies a set of two coupled Hamilton equations, $i\,\partial\hat{\varphi}(z,t)/\partial t=ic^2\,\hat{\Pi}(z,t)$ and $i\,\partial\hat{\Pi}(z,t)/\partial t=-i\,[m^2c^4-c^2(\partial/\partial z)^2]/c^2\,\hat{\varphi}(z,t)$.

With regard to the time-evolution discussed below, it is important to point out that $\hat{H}$ is local only with respect to the operator $\hat{\varphi}(z)$, while when expressed in terms of $\hat{a}(z)$ even its interaction-free part $\hat{H}_0$ is non-local. If created by $\hat{a}^\dagger(z)$, the properties of a particle at z can be influenced instantaneously by particles at other locations z'. This finding is also consistent with the fact that $[\hat{\varphi}(z=0,t=0),\ \hat{\varphi}(z,t)]=0$, while $[\hat{a}(z=0,t=0),\ \hat{a}(z,t)]\neq 0$ outside the light cone, $t<c|z|$.

## 3. The two position yardsticks based on $\hat{\varphi}(z)$ and $\hat{a}(z)$

In order to visualize the dynamics as predicted by $\hat{H}$, we need to associate a spatial density with the state $|\psi(t)\rangle$. In contrast to the corresponding momentum density $\langle\hat{a}^\dagger(p)\,\hat{a}(p)\rangle$, this association is non-trivial and (at least) two yardsticks have been proposed to extract position dependent information from $|\psi(t)\rangle$. Two operators can be used to create a particle at "location z" from the vacuum state $|vac\rangle$. The first one is the field operator $\hat{\varphi}(z)$ and one can find statements in numerous standard textbooks [14-16] stating that it creates a particle located at position z, $\hat{\varphi}^\dagger(z)|vac\rangle$. The second one is the Fourier transform of the momentum mode operator $\hat{a}(z)\equiv(2\pi)^{-1/2}\int dp\,\hat{a}(p)\exp(ipz)$ (as introduced above), leading to $\hat{a}^\dagger(z)|vac\rangle$. In quantum optics $\hat{a}(z)$ is called the positive frequency operator associated with the photon intensity [17]. It is also the Newton-Wigner field [4,12,13] for bosonic systems and (similar to the momentum operators) $\hat{a}(z)$ fulfills the equal-time commutation relationship $[\hat{a}(z_1),\ \hat{a}^\dagger(z_2)]_-=\delta(z_1-z_2)$. Analogous to $\hat{a}(k)$, which is interpreted as the operator creating a particle with fixed momentum k, the operator $\hat{a}(z)$ could be interpreted as the creation operator for the position mode located at z. We also note that for

any state there is the Parseval-like equality $\int dp \, \langle \hat{a}^\dagger(p) \, \hat{a}(p) \rangle = \int dz \, \langle \hat{a}^\dagger(z) \, \hat{a}(z) \rangle$, which helps us to interpret the data in terms of particles.

The simple definitions for position states as $\hat{a}^\dagger(z)|vac\rangle$ or $\hat{\varphi}^\dagger(z)|vac\rangle$ lead to an infinite normalization of the corresponding states, which is not so convenient for numerical purposes. We therefore define in this work the position states in a slightly more complicated way as the limit $\Delta \to 0$ of $s_\Delta(z)$, where $\int dz \, |s_\Delta(z)|^2 = 1$

$$|z;\hat{a}\rangle \equiv \lim_{\Delta \to 0} \int dz' \, s_\Delta(z-z') \, \hat{a}^\dagger(z') \, |vac\rangle \tag{3.1}$$

$$|z;\hat{\varphi}\rangle \equiv \lim_{\Delta \to 0} (2m)^{1/2} \int dz' \, s_\Delta(z-z') \, \hat{\varphi}^\dagger(z') \, |vac\rangle \tag{3.2}$$

In the second definition we have arbitrarily included the factor $(2m)^{1/2}$ to guarantee that both states have the same nonrelativistic limit ($c \to \infty$). In the zero-width limit $\Delta \to 0$, the function $s_\Delta(z)$ approaches the square root of the Dirac delta function, $s_\Delta(z)^2 \to \delta(z)$. For numerical realizations of $s_\Delta(z)$ we have used $s_\Delta(z) = (2/\pi)^{1/4} \Delta^{-1/2} \exp[-(z/\Delta)^2]$.

It is important to note that two different states $|z;\hat{a}\rangle$ are orthogonal to each other, $\langle z_1;\hat{a}|z_2;\hat{a}\rangle = 0$ for $z_1 \neq z_2$, while the states $|z;\hat{\varphi}\rangle$ are *not* and therefore they cannot be viewed as eigenstates of any hermitian position operator. Using the above definitions, one can show that the two yardsticks are related to each other via a *non-local* but linear transformation

$$|z;\hat{\varphi}\rangle = \int dz' \, L_{1/2}(z-z') \, |z';\hat{a}\rangle \tag{3.3a}$$

$$|z;\hat{a}\rangle = \int dz' \, 2 \, I_{1/2}(z-z') \, |z';\hat{\varphi}\rangle \tag{3.3b}$$

where the functions in the integral were defined in Eqs. (2.5). Note that the two functions also permit us to relate the operators to each other, via

$$\hat{\varphi}(z) = \int dz' \, L_{1/2}(z-z') \, [\hat{a}(z') + \hat{a}^\dagger(z')] \tag{3.4a}$$

$$\hat{\Pi}(z) \;=\; -i \int dz' \; I_{1/2}(z\text{-}z') \; [\hat{a}(z') - \hat{a}^{\dagger}(z')] \tag{3.4b}$$

$$\hat{a}(z) \;=\; \int dz' \; [I_{1/2}(z\text{-}z') \hat{\varphi}(z') + i \, L_{-1/2}(z\text{-}z') \hat{\Pi}(z')] \tag{3.4c}$$

If we define the position distribution for a state $|\psi\rangle$ via the expectation value of the spatial occupation number given by the corresponding operator product, we find

$$\rho_{\hat{a}}(z) \;\equiv\; \langle\psi| \, \hat{a}^{\dagger}(z)\hat{a}(z) \, |\psi\rangle \;=\; |\langle z;\hat{a}|\psi\rangle|^2 \tag{3.5a}$$

$$\rho_{\hat{\varphi}}(z) \;\equiv\; \langle\psi| \, \hat{\varphi}^{\dagger}(z)\hat{\varphi}(z) \, |\psi\rangle \, /m \;=\; |\langle z;\hat{\varphi}|\psi\rangle|^2 \tag{3.5b}$$

Note that the second equalities only holds if $|\psi\rangle$ describes a single particle. If as a special case the state is chosen to be $|\psi\rangle=|z_1;\hat{a}\rangle$, we find consistently $\rho_{\hat{a}}(z)=\delta(z\text{-}z_1)$, while for the state $|\psi\rangle=|z_1;\hat{\varphi}\rangle$, neither $\rho_{\hat{a}}(z)$ nor $\rho_{\hat{\varphi}}(z)$ are localized. We also note that the two corresponding complex wave functions for a single particle state $|\psi\rangle$ can be related to each other via $\langle z;\hat{a}|\psi\rangle = \int dz' \, 2 \, I_{1/2}(z\text{-}z') \langle z';\hat{\varphi}|\psi\rangle$ and $\langle z;\hat{\varphi}|\psi\rangle = \int dz' \, L_{-1/2}(z\text{-}z') \langle z';\hat{a}|\psi\rangle$, respectively. The fact that $L_{1/2}(z)$ is positive shows that the "spatial amplitude" for any single particle state $\langle z';\hat{\varphi}|\psi\rangle$ in z is in general wider than for $\langle z';\hat{a}|\psi\rangle$.

## 4. Time evolution of the densities for free particles ($\lambda$=0)

Let us first analyze the time-evolution of the same initial state $|\psi(t)\rangle$ under the force-free Hamiltonian Eq. (2.3a), but viewed under the two position yardsticks $\rho_{\hat{a}}(z,t)$ and $\rho_{\hat{\varphi}}(z,t)$. We choose as the initial state $|\psi(t{=}0)\rangle \equiv \int dz \, G(z) \, \hat{a}^{\dagger}(z)|vac\rangle$, where $G(z)$ is the corresponding quantum mechanical wave function, such that its initial density $\rho_{\hat{a}}(z,t{=}0)$ is simply $|G(z)|^2$. The time evolution is given by $|\psi(t)\rangle = \int dp \, G(p) \exp[-i\omega(p)t] \, \hat{a}^{\dagger}(p)|vac\rangle$, where $G(p)$ denotes the Fourier transform $(2\pi)^{-1/2}\int dz \, G(z) \exp[-ipz]$.

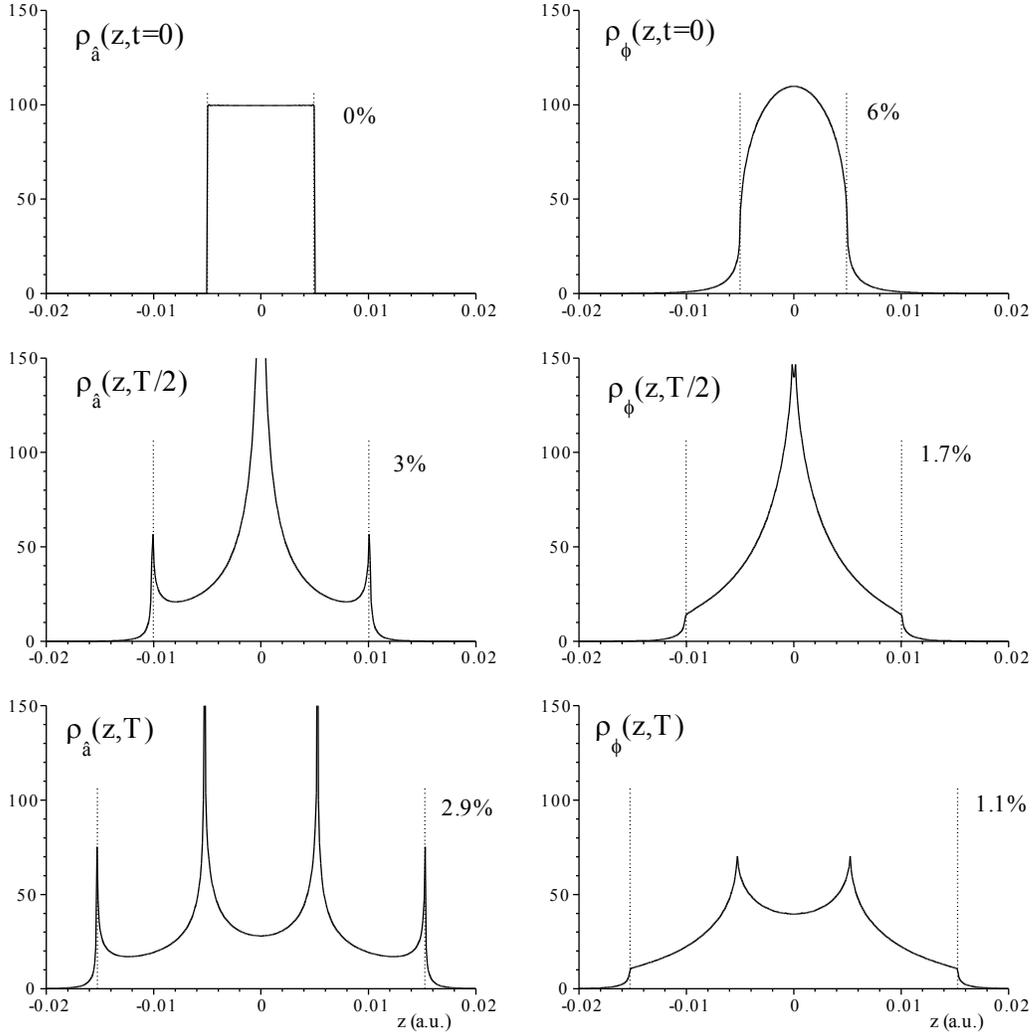

**Figure 1**

The initial and the time evolved spatial densities $\rho(z,T/2)$ and $\rho(z,T)$ for the same quantum state $|\Psi(t)\rangle$ computed using the $\hat{a}$- and $\hat{\varphi}$-based yardsticks. For comparison, the two vertical dashed lines indicate the light cones at $z=\pm(w+ct)$ and the percentage is the fraction of the density outside of both light cones [$w=0.005$ a.u., $T=7.5\times10^{-5}$ a.u.]

For the data displayed in Figure 1, we have assumed that the amplitude $G(z)$ is nonzero only for $|z|<w$, i.e. $G(z) = (2w)^{1/2}\,\theta(w-|z|)$, where $\theta(\ldots)$ denotes the Heaviside unit step function, defined as $\theta(z) \equiv (1+|z|/z)/2$ and $2w$ is the width of the initial state. The graphs in the left column show the Newton-Wigner presentation of the spatial density $\rho_{\hat{a}}(z,t)$ and the right column is the distribution $\rho_\varphi(z,t)$ defined in Eqs. (3.5). For better comparison, the latter was normalized to $\int dz \, \rho_\varphi(z) = 1$, whereas $\rho_{\hat{a}}(z)$ automatically fulfills $\int dz \, \rho_{\hat{a}}(z) = 1$.

The upper row in Figure 1 shows the two initial distributions. While $\rho_{\hat{a}}(z)$ is sharply localized between $-w < z < w$, the yardstick based on $|z; \hat{\varphi}\rangle$ suggests that the distribution $\rho_\varphi(z)$ is infinitely extended. This is consistent with the properties of the integration kernel $I_{-\frac{1}{2}}$ discussed above. We have not been able to construct any normalizable single particle state $|\psi\rangle$ such that its spatial density $\rho_\varphi(z)$ has a compact spatial support. This feature makes it more difficult to define unambiguously the corresponding light cone as a gauge to quantify a possible superluminal component [18] of $\rho_\varphi(z)$.

While for the small spatial widths $w < 1/c$ in the Figure the two distributions $\rho_{\hat{a}}(z)$ and $\rho_\varphi(z)$ are rather different, for larger widths they become more similar to each other. For states that contain only small momentum contributions (corresponding to a large spatial width w) we have $\rho_{\hat{a}}(z) \approx \rho_\varphi(z)$ under the appropriate normalization. This is consistent as the difference between the two position yardsticks is purely a relativistic effect and in the limit $c \to \infty$ the field in Eq. (2.2a) turns into $\hat{\varphi}(z) \to (2m)^{-1/2} [\hat{a}(z) + \hat{a}^\dagger(z)]$.

The middle row shows the distributions at a later time. The dashed vertical reference lines mark the locations $\pm(w+ct)$ evolving with speed c. This permits us to evaluate the portions of the distributions that are outside the light cone. We see that about 3% of the distribution $\rho_{\hat{a}}(z)$ has moved outside the light cone, suggesting a superluminal spreading. Refs. [19,20] have analyzed this portion more systematically and showed that for longer times this portion reduces to zero such that this superluminal effect is transient.

For comparison we have also computed the portion of the distribution that is outside of the light cone for $\rho_\varphi(z)$. Here this portion shrinks from 6% (characteristic of the initially extended distribution) to zero. Quite interestingly, the density develops rather sharp boundaries along the borderline of the two light cones to the left and to the right.

## 5. Time evolution for two interacting particles ($\lambda \neq 0$)

The general question of whether an interaction between two particles is instantaneous or retarded is extremely difficult to examine. We consider here only the special case of the $\hat{\int}^4$ system, which describes only one type of indistinguishable particles. Furthermore, as this Hamiltonian is local in $z$, the interaction is short ranged and therefore mainly confined to regions where the densities of the particles overlap in $z$. As the densities $\rho_\varphi(z)$ evolve subluminally, it is therefore reasonable to assume that the interaction does well. However, the $|z;\hat{\int}\rangle$ based yardstick does not allow for initially localized distributions, which makes it difficult to assign portions to only one particle and to identify the effect of one particle on the other.

The propagation with respect to the $|z;\hat{a}\rangle$-based yardstick, however, is superluminal and therefore could have the potential of permitting an almost instant communication between two distant particles. As already the free Hamiltonian Eq. (2.3a) when expressed in terms of the complete set of operators $\hat{a}(z)$ is non-local, as $V(z_1,z_2)\neq 0$ for $z_1\neq z_2$, two initially localized and separate particles could interact even if their spatial densities do not overlap. We describe some first steps towards an investigation whether the presence of one particle affect the time evolution of the spatial density of the other particle in space-like regions. We are not providing an ultimate answer, but rather some first suggestions to obtain a little insight into this quite difficult question.

We have prepared the initial state as $|\psi(t=0)\rangle \equiv \iint dz_1 dz_2 \, G(z_1-x) \, G(z_2-y) \, |z_1;\hat{a}\rangle \, |z_2;\hat{a}\rangle$, corresponding to two particles that are initially centered around $z=x$ and $z=y$ according to the Newton-Wigner yardstick. Here and below we assume that x and y are initially sufficiently far apart (or equivalently $G(z)$ is sufficiently narrow) so that the spatial overlap of the two initial wave functions can be neglected, leading to a sum of two disjoint densities $\rho_{\hat{a}}(z) = |G(z-x)|^2 + |G(z-y)|^2$. We are interested again in space-like regions, such that the time t has to be less than it takes for a light pulse to travel from one particle to another, $t<|z_1-z_2|/c$. The key question is whether the time evolved density $\rho_{\hat{a}}(z,t)$ remains just the sum of the individual densities, or whether the densities spread asymmetrically, as a possible manifestation on an interaction.

We have computed the evolution of the density for short times, such that $\exp(-i\hat{H}t)$ can be approximated by $1-i\hat{H}t-(\hat{H}t)^2/2$.

$$\rho_{\hat{a}}(z,t) \equiv \langle\psi(t=0)|\hat{a}^\dagger(z,t)\,\hat{a}(z,t)|\psi(t=0)\rangle$$

$$\approx \langle\psi(t=0)|\left(1+i\hat{H}t-\hat{H}^2t^2/2\right)\hat{a}^\dagger(z)\hat{a}(z)\left(1-i\hat{H}t-\hat{H}^2t^2/2\right)|\psi(t=0)\rangle \qquad (5.1)$$

This short time expansion warrants two comments. First, as is generic to any non-unitary time evolution, the norm of the state and the corresponding density are not necessarily conserved. Second, the energy spectrum of the initial state determines the temporal range of validity. Spatially very narrow states contain high-momentum components which limit the maximum value of the time. For example, the validity of the expansion for states with compact support is not clear. In the opposite limit for a state with vanishing momentum, however, the short time expansion is (trivially) valid for all times t.

We obtain the constant term $\rho_{\hat{a}}(z,t=0)$, a term that is linear in time, $it\langle[\hat{H},\hat{a}^\dagger(z)\hat{a}(z)]\rangle$, and three terms that are quadratic in time, $t^2\{\langle\hat{H}\hat{a}^\dagger(z)\hat{a}(z)\hat{H}\rangle - \langle\hat{H}^2\hat{a}^\dagger(z)\hat{a}(z)\rangle - \langle\hat{a}^\dagger(z)\hat{a}(z)\hat{H}^2\rangle\}$ and neglect for consistency the higher order terms in time. If we decompose the Hamiltonian $\hat{H}=\hat{H}_0+\hat{V}$ of Eq. (2.1) into the free and interacting parts and multiply the operators out in Eq. (5.1), we can find some numerically tractable expressions for these terms. As the derivations are cumbersome and the final expressions are rather lengthy, we refer the reader to the Appendix A for more details. We therefore present the results graphically here. In the Appendix we show that the linear terms vanish such that only the quadratic terms contribute. As we are only interested in the leading order of the coupling constant we find

$$\rho_{\hat{a}}(z,t) = \rho_{\hat{a}}(z,t=0) + r_{free}(z)t^2 + \lambda\, r_{int}(z)t^2 + O(t^3) \qquad (5.2)$$

where the interaction-free ($\lambda=0$) part $r_{free}(z) \equiv \langle\hat{H}_0\hat{a}^\dagger(z)\hat{a}(z)\hat{H}_0\rangle - \langle\hat{H}_0^2\hat{a}^\dagger(z)\hat{a}(z)\rangle/2 - \langle\hat{a}^\dagger(z)\hat{a}(z)\hat{H}_0^2\rangle/2$ describes the evolution of both particles independent of each other.

The more important part for our discussion is the term linear in the coupling constant, $\lambda^{-1} r_{int}(z) \equiv \langle\psi(t=0)|\hat{H}_0\hat{a}^\dagger(z)\hat{a}(z)\hat{V}|\psi(t=0)\rangle - \langle\psi(t=0)|(\hat{V}\hat{H}_0+\hat{H}_0\hat{V})\hat{a}^\dagger(z)\hat{a}(z)|\psi(t=0)\rangle/2 + c.c.$, as it describes how the $\hat{j}^4$-interaction affects the dynamics of each particle. The evaluation of this term involves 15-fold integrals that can be reduced to a slighly less complicated form whose expression we derive in Appendix A. We just focus here on its graphical presentation in Figure 2. Here the initial amplitudes

G(z+0.02) and G(z-0.02) were chosen as very narrow Gaussians with a width w that is smaller than the spacing |x-y|=0.04 by about a factor of ten.  While the correction terms $r_{free}(z)$ (not shown) are symmetric around z=x and z=y and reflect an independent time evolution, we see that $r_{int}(z)$ corrects the density in an asymmetric way.  We have shown the data for two different initial widths w to examine whether the asymmetry could be simply a consequence of the (unavoidable) initial overlap of G(z-x) and G(z-y).  For z=0 the ratio of the initial densities $\rho_{\hat{a}}(z=0,t=0)$ for w=0.0025 and w=0.005, respectively, is practically zero, due the rapid Gaussian fall-off.  The corresponding ratio for the terms $r_{int}$ at z=0, however, is about one sixth.  This comparison suggests that the cause of the asymmetric form of the correction term $r_{int}(z)$ around z=±0.02 should be of a kinematic nature and not simply a consequence of the asymmetry asscociated with the initial overlap.

As this correction term $r_{int}(z)$ is linear in λ (in contrast to many other quantum field theoretical interactions where the resulting forces scale quadratically in the coupling strength and are therefore either repulsive or attractive), the direction of the force between the particles for the $\hat{j}^4$-system seems to depend on λ.  For our choice of a positive sign of λ, we find that the probability density due to the interaction is increased between both particles, as $r_{int}(z)$ is mostly positive in that region, suggesting possibly an attractive force.  We also see that the positions of the two minima of $r_{int}(z)$ are shifted inwards.  This drift is especially visible for the larger width w=0.005 a.u.  Certainly more studies on the details of this interaction beyond the main theme of this work would be quite interesting.  For first work in this direction we refer the reader to a recent publication [21].

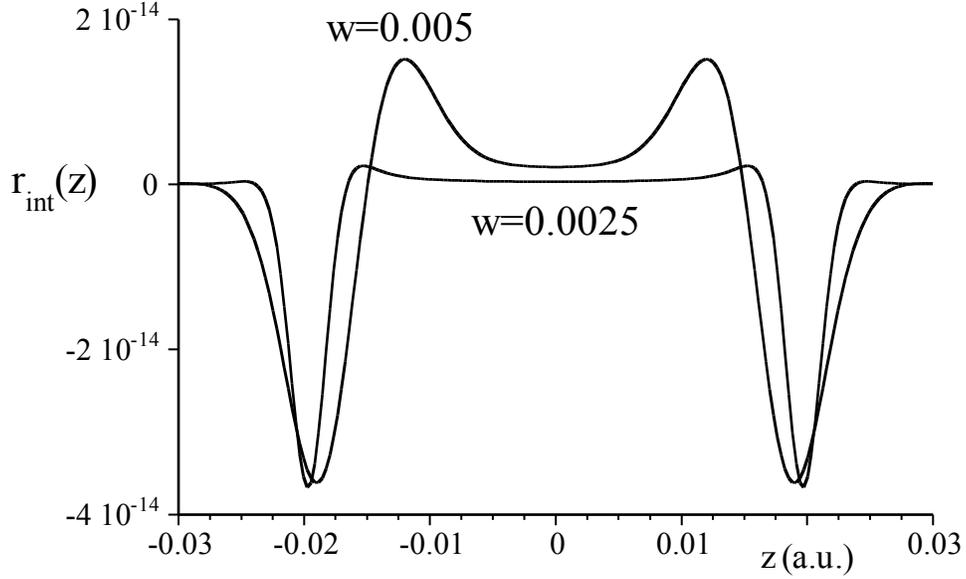

**Figure 2** The additive correction to the spatial density $r_{int}(z)$ that is associated with the interaction. It is shown for two different initial widths w, where $G(z)=(2w^2/\pi)^{-1/4} \exp(-z^2/w^2)$.

## 6. Transformation properties of the yardsticks for moving frames

In this section we examine how the mathematical expressions for the observables associated with the two yardsticks need be to modified when viewed from a coordinate system that moves with positive velocity v relative to the original reference frame. For simplicity we introduce here the rapidity parameter $\theta=\tanh(v/c)$. In Appendices B and C we give more details about the properties of the corresponding boost operator $\exp[iKc\theta]$ (abbreviated by $\hat{B}$) that transforms any operator $\hat{A}$ into the moving frame according to $\hat{A}(\theta)=\hat{B}^\dagger\hat{A}\hat{B}$.

To simplify our notation, we assume that the two yardstick states evolve in time according to $\exp(-i\hat{H}t)\,|z\rangle$, which we abbreviate as $|z,t\rangle$. The system is described from the moving frame as $|\Psi;\theta\rangle \equiv \hat{B}|\Psi\rangle$ to guarantee that $\langle\Psi|\hat{A}(\theta)|\Psi\rangle = \langle\Psi;\theta|\hat{A}|\Psi;\theta\rangle$. The corresponding yardstick states, however, need to be transformed as $|z;\theta\rangle = \hat{B}^\dagger|z\rangle$ to guarantee that $|\langle z;\theta|\Psi\rangle|^2$ is the density as seen by the moving observer for the state described in the original frame as $|\Psi\rangle$ with density $|\langle z|\Psi\rangle|^2$. Since we are transforming here the yardsticks rather than the state this corresponds to the Heisenberg representation.

More specifically, the transformation of the $\hat{\varphi}$-based yardstick into the moving frame leads to $\hat{B}^{\dagger}|z,t;\hat{\varphi}\rangle = \hat{B}^{\dagger}\,\hat{\varphi}(z,t)\,|vac\rangle$.  If we insert the unit operator $\hat{B}\,\hat{B}^{\dagger}$ before the vacuum state and use the invariance of $|vac\rangle$, we obtain $\hat{B}^{\dagger}|z,t;\hat{\varphi}\rangle = \hat{B}^{\dagger}\,\hat{\varphi}(z,t)\,\hat{B}\,|vac\rangle$.  In Appendix C we have shown that $\hat{B}^{\dagger}\,\hat{\varphi}(z,t)\,\hat{B} = \hat{\varphi}(z_{-\theta},t_{-\theta})$, so therefore $\hat{B}^{\dagger}|z,t;\hat{\varphi}\rangle = \hat{\varphi}(z_{-\theta},t_{-\theta})\,|vac\rangle = |z_{-\theta},t_{-\theta};\hat{\varphi}\rangle$, where the pair $(z_{-\theta},t_{-\theta})$ is just the usual Lorentz transformed variables $(z_{-\theta},t_{-\theta}) = L_{-\theta}(z,t)$.  Here the two-component vector is defined as $L_{\theta}(a,b) \equiv [a\,Cosh\theta - b\;c\;Sinh\theta,\;b\,Cosh\theta - a/c\;Sinh\theta]$.  In other words, for any single-particle state the expansion amplitude with respect to the $\hat{\varphi}$-based yardstick transforms according to the usual Lorentz equations

$$\langle\psi|\hat{B}^{\dagger}|z,t;\hat{\varphi}\rangle = \langle\psi|z_{-\theta},t_{-\theta};\hat{\varphi}\rangle \qquad (6.1)$$

The corresponding transformation for the $\hat{a}$-based yardstick basis states is slightly more complicated [19] as the transformation of $\hat{a}(z,t)$ cannot be simply reduced to a simple operation on its arguments z and t.  In fact, we derive in Eq. (C2a) the transformation law $\hat{B}^{\dagger}\,\hat{a}(z)\,\hat{B} = \int dz'$ $F_{\theta}(z_{-\theta} - z',t_{-\theta})\,\hat{a}(z')$.  As a result, the wave function transforms as

$$\langle\psi|\hat{B}^{\dagger}|z,t;\hat{a}\rangle = \int dz'\; F_{\theta}(z_{-\theta} - z',t_{-\theta})\,\langle\psi|z';\hat{a}\rangle \qquad (6.2)$$

where the integration kernel is given by

$$F_{\theta}(z_{-\theta} - z',t_{-\theta}) \;\equiv\; (2\pi)^{-1}\!\int dq\;[\omega(p)/\omega(q)]^{1/2}\,exp[-i\omega(q)t_{-\theta}+iq(z_{-\theta}-z')] \qquad (6.3)$$

If we set t=0 in Eq. (6.2) we obtain equivalently $\langle\psi|\hat{B}^{\dagger}|z;\hat{a}\rangle = \int dz'\,f_{\theta}(z,z')\,\langle\psi|z';\hat{a}\rangle$, where $f_{\theta}(z,z') \equiv (2\pi)^{-1}\!\int dp\;[\omega(q)/\omega(q)]^{1/2}\,exp[ipz-ip_{-\theta}z']$.  Note that this function has the interesting symmetry property $f_{\theta}(z,z') = f_{-\theta}(z',z)^{*}$ and $\int dz'\,f_{\theta}(z,z')\,f_{-\theta}(z',z'') = \delta(z-z'')$.

In Figure 3 we have graphed the corresponding boost-transformed density $\rho_{\hat{a}}(z;\theta)$.  It is clear that even for a special state for which the initial density is localized for the $\hat{a}$-based yardstick, any other frame predicts an infinitely extended density.  In other words, a state with compact spatial support is a rather unique special case even within the â-based yardstick.  In order to quantify the importance of the correct transformation law, we have also computed the density had we applied the usual Lorentz formula (which is incorrect for the â-based yardstick).  We note that the two transformed densities are not identical but qualitatively rather similar.  The boost transformation is unitary and leaves the norm of the state $\langle \Psi | \Psi \rangle$ unchanged.  However, we point out that only the norm $\int dz\, \rho_{\hat{a}}(z,t)$ is conserved under the boost, whereas $\int dz\, \rho_{\phi}(z,t)$ is not.  This is directly related to the fact that in the single particle space $\int dz\, |z;\hat{a}\rangle\langle z;\hat{a}|$ is the unit operator but $\int dz\, |z;\hat{\varphi}\rangle\langle z;\hat{\varphi}|$ is not (due to the lack of orthogonality).

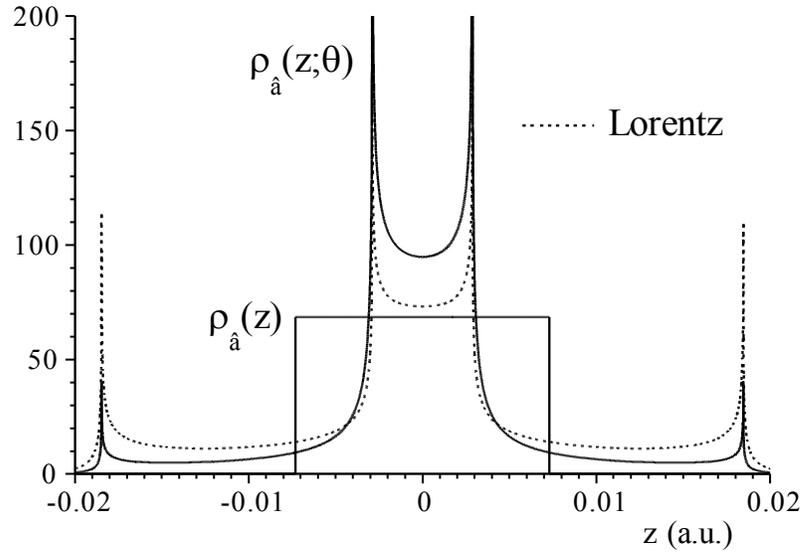

**Figure 3**

The density $\rho_{\hat{a}}(z)$ of a state $|\Psi\rangle$ together with the density $\rho_{\hat{a}}(z;\theta)$ an observer would see in coordinate frame that moves with velocity v=100 corresponding to $\theta$=0.93.  For comparison, the dashed line shows the (normalized) density one obtained using (incorrectly) the Lorentz transformation formula.  The width of the density in the rest frame is w=7.3×10$^{-3}$.

The strong similarity of the time- and velocity-translated densities of Figs. 1 and 3, respectively, is worth noting.  The four–peak structure of the time-translated densities in Figs. 1 was associated with the sharp edges of the initial density $\rho_{\hat{a}}(z,t=0)$ representing regions of very large velocities.  At later times each edge breaks into two peaks that propagate with the speed of light c.

As a result, the outer peaks are located at z=±(w+ct), and the inner two peaks at z=±(w−ct). A similar four-peak structure arises if the initial density $\rho_{\hat{a}}(z,t=0)$ is seen from an moving frame.

An estimate of the locations of these characteristic markers of the density can be easily found. As the initial state was chosen real, the time-reversal symmetry predicts that $\rho_{\hat{a}}(z,-t)=\rho_{\hat{a}}(z,t)$. In other words, the location of the right most peak (moving with c) evolves in time as $z_4(t)=w+c|t|$. If we use the usual Lorentz formulas to predict the location $z'_4$ where the "event" [$z_4(t),t$] would be observed in a moving frame (at time t'=0), we have to compute $z'_4 = z_4(t)$ Cosh$\theta$ − t c Sinh$\theta$. The time in the moving frame t'= t Cosh$\theta$ − $z_4(t)/c$ Sinh$\theta$ has to be equated to zero to find the corresponding moment in time in the original frame. We obtain for this time t=w/c Sinh$\theta$/(Cosh$\theta$−Sinh$\theta$). If we insert this term into the equation for $z'_4$, we obtain $z'_4 = w$ Exp[$\theta$]. This expression predicts the location of the right most peak $z'_4=1.85\times10^{-2}$ a.u. for the moving observers v=100 as shown in the Figure.

More general, if the original density $\rho_{\hat{a}}(z)$ is nonzero and constant between $z_L$ and $z_R$, the four peaks characteristic of the boosted density would occur at locations $z_L$Exp[$\theta$], $z_L$Exp[−$\theta$], $z_R$Exp[−$\theta$] and $z_R$Exp[$\theta$]. These multiplicative factors are interesting and illustrate the fact that while the original density is symmetric around its center $(z_R-z_L)/2$, the boosted one does not have any symmetry as the separation $z_{L,R}$2Sinh$\theta$ between the two peaks associated with each edge depends on the location of the edge. Furthermore, the locations of two peaks approach z=0 for large rapidity $\theta$. The same conclusion can be also obtained by the appropriate projections in a Minkowski diagram.

## 7. Brief discussion and outlook

Using concrete numerical calculations we have illustrated the predictions of two proposals to assign a spatial probability distribution to the same quantum field theoretical state for a single particle. The distributions associated with $\hat{\varphi}(z)$ are in general wider than the ones based on the operator $\hat{a}(z)$. Furthermore, in contrast to $\hat{\varphi}(z)$, the operator $\hat{a}(z)$ permits localized densities with compact support whose time-evolution reveals a transient superluminal propagation. While the possibility of localized states is essential from a conceptual point of view to define mutually orthogonal position eigenstates, compact support is also a rather unique property as any state becomes delocalized if viewed from any velocity or time shifted coordinate frame.

Unfortunetely, both yardsticks have properties that could cause some concern. The $\hat{\varphi}$-based yardstick cannot generate states that are mutually orthogonal with each other, which is a necessary feature for eigenstates of a position operator. The transformation properties of the wave functions associated with the â-based yardstick under boosts are different from the usual (Lorentz transformation based) covariant scheme. It is important to point out [23] that covariance is not a condition for the physical validity of any operator, but a technical simplification when computing the functional form seen from a moving coordinate frame. For example, the momentum creation operator â(p) does not have this (covariance) property. After all, the underlying dynamics fulfills the Poincare relationships and is therefore relativistically invariant as required. In fact, the Newton-Wigner operator can be generalized to become covariant; see the works by Fleming [12,13].

The observed superluminal propagation of a wave packet would constitute a serious problem for the â-based yardstick if one could show that there is a moving frame in which cause and effect would be observed to be reversed and therefore violate the principle of causality. However, the usual Lorentz formulas (on which arguments for the reversal of cause and effect are usually based on) do not describe the correct transformation for this yardstick as we have discussed.

An important question concerns the physical validity of the two yardsticks. To the best of our knowledge, it is presently not clear which one of them describes the actual position of a physical detector. In this work we have used the bare vacuum, bare annihilation and creation operators as well as the free field operator as tools for defining localized particle states. It is important to understand how these definitions are affected by the presence of interactions. One possible solution would be to use dressed particle operators introduced by Greenberg and Schweber [24]. However, in this case the position operator and the notion of localization become dependent on the interaction strength, which is not desirable. An alternative approach is to apply the unitary dressing transformation directly to the Hamiltonian, so that definitions of particles and their observables do not depend on interactions (see section 10.2 in [29]).

In Section 5 we showed that the superluminal propagation (discussed in Section 4) can evolve in an asymmetric way, possibly suggesting an almost instantaneous interaction between two particles. However, this issue is much more complicated and far from resolved. One could also take the view point that our chosen initial state at t=0 does not really correspond to the true birth moment when both particles where created but it is just a particular temporal snapshot of a system that describes two particles that have already been interacting with each other for t<0. As a result, the computed dynamics for t>0 would be just a continuation of the past interaction and one should not conclude that each dynamical effect observed for t>0 has no cause at t<0. Furthermore, the assumption of the absence of any interaction for t<0 or the assumption of creating two particles out of the vacuum at t=0 would require a time-dependent Hamiltonian, which would invalidate our Poincare-group based approach. Within this view point it is also difficult to define at all what a retardation would mean, as a precise reference point in time is difficult to identify.

In addition to these conceptual difficulties, there are also purely technical issues that need to be addressed in future work. Our preliminary findings were based on a short-time expansion of the time evolution propagator, whose validity is nontrivial when high momenta, which are characteristic of densities with compact support, are involved. We also note that even in the limit of vanishing coupling $\lambda$ the density could contain small degrees of asymmetry that are associated with the intereference that is expected when the densities of the two particles overlap.

Furthermore, the $\hat{\varphi}^4$-coupling can increase the number of bare particles and a non-pertubative calculation would require us to begin the evolution with two dressed states. Due to numerical constraints and also to be consistent with a perturbative approach that is linear in $\lambda$, the initial state in Section 5 had to be chosen as two bare particles. To include the dressing of a particle would require a significantly larger Hilbert space [25], but it seems to be very worthwhile to address this in a future work. First attempts to define dressed operators can be found in Refs. [24,26,27]. It is our hope that this work can trigger more interest and studies on the temporal characteristics of quantum field theoretical interactions.

## Acknowledgements

We enjoyed several helpful discussions with Profs. C.C. Gerry, C. Müller and Y.T. Li. QS and RG acknowledge the kind hospitality of their host institutions during their sabbatical leaves. This work has been supported by the NSF and the NSFC (#11128409).

**Appendix A**

Here we derive the analytical expression for the time evolved spatial density in lowest order perturbation theory in the coupling constant $\lambda$. The initial state is given by

$$|\psi(t=0)\rangle \equiv \iint dz_1 dz_2\ G(z_1\text{-}x)\ G(z_2\text{-}y)\ |z_1;\hat{a}\rangle\ |z_2;\hat{a}\rangle \tag{A1}$$

corresponding to two particles that are initially centered around $z=x$ and $z=y$. If we assume that the overlap between the two spatial amplitudes is negligible, i.e. $\int dz\ G(z\text{-}x)G(z\text{-}y)\approx 0$, the density $\rho_{\hat{a}}(z,t=0) \equiv \langle\psi(t=0)|\hat{a}^{\dagger}(z)\hat{a}(z)|\psi(t=0)\rangle = |G(z\text{-}x)|^2+|G(z\text{-}y)|^2$. For the time evolution we obtain

$$\rho_{\hat{a}}(z,t) \equiv \langle\psi(t=0)|\hat{a}^{\dagger}(z,t)\hat{a}(z,t)|\psi(t=0)\rangle$$

$$\approx \langle\psi(t=0)|\left(1+i(\hat{H}_0+\hat{V})t-(\hat{H}_0+\hat{V})^2 t^2/2\right)\hat{a}^{\dagger}(z)\hat{a}(z)\left(1-i(\hat{H}_0+\hat{V})t-(\hat{H}_0+\hat{V})^2 t^2/2\right)|\psi(t=0)\rangle$$

$$\equiv \rho_{\hat{a}}(z,t=0)\ +\ t\ r_{\text{free}}^{(1)}(z)\ +\ t^2\ r_{\text{free}}^{(2)}(z)\ +\ t\,\lambda\ r_{\text{int}}^{(1)}(z)\ +\ t^2\,\lambda\ r_{\text{int}}^{(2)}(z)\ +\ O(t^3,\lambda^2)$$

$$\tag{A2}$$

where we neglect the quadratic terms in the coupling constant. The lowest-order terms are defined as

$$r_{\text{free}}^{(1)}(z) \equiv i\ \langle[\hat{H}_0, \hat{a}^{\dagger}(z)\hat{a}(z)]_-\rangle \tag{A3a}$$

$$r_{\text{int}}^{(1)}(z) \equiv \lambda^{-1}\ i\ \langle[\hat{V}, \hat{a}^{\dagger}(z)\hat{a}(z)]_-\rangle \tag{A3b}$$

$$r_{\text{free}}^{(2)}(z) \equiv -\langle\hat{H}_0^2\ \hat{a}^{\dagger}(z)\hat{a}(z)\rangle/2 - \langle\hat{a}^{\dagger}(z)\hat{a}(z)\ \hat{H}_0^2\rangle/2 + \langle\hat{H}_0\ \hat{a}^{\dagger}(z)\hat{a}(z)\hat{H}_0\rangle \tag{A3c}$$

$$r_{\text{int}}^{(2)}(z) \equiv \lambda^{-1}\ \{\langle\hat{H}_0\ \hat{a}^{\dagger}(z)\hat{a}(z)\ \hat{V}\rangle + \langle\hat{V}\ \hat{a}^{\dagger}(z)\hat{a}(z)\ \hat{H}_0\rangle$$

$$-\langle(\hat{V}\hat{H}_0+\hat{H}_0\hat{V})\ \hat{a}^{\dagger}(z)\hat{a}(z)\rangle/2 - \langle\hat{a}^{\dagger}(z)\hat{a}(z)(\hat{V}\hat{H}_0+\hat{H}_0\hat{V})\rangle/2\ \} \tag{A3d}$$

The two terms that are linear in time, $r_{\text{free}}^{(1)}(z)$ and $r_{\text{int}}^{(1)}(z)$, can be shown to vanish if one uses $[\hat{a}^{\dagger}(z_1)\hat{a}(z_2), \hat{a}^{\dagger}(z)\hat{a}(z)]= \hat{a}^{\dagger}(z_1)\hat{a}(z_2)[\delta(z\text{-}z_2)\text{-}\delta(z\text{-}z_1)]$. Among the quadratic terms we focus here only on the expectation values $\langle\hat{H}_0\ \hat{a}^{\dagger}(z)\hat{a}(z)\ \hat{V}\rangle + \langle(\hat{V}\hat{H}_0+\hat{H}_0\hat{V})\hat{a}^{\dagger}(z)\hat{a}(z)\rangle/2$ + c.c. They are the most important ones for our discussion, as they are linear in the coupling constant λ. We need to simplify this expression for $r_{int}(z)$ to make it accessible to numerical analysis.

Let us begin with the term $\langle \hat{H}_0 \, \hat{a}^\dagger(z)\hat{a}(z)\,\hat{V}\rangle$. If we insert the relevant non-vanishing parts of Eqs. (2.3) for $\hat{H}_0$ and $\hat{V}$ into Eq. (A3) we obtain the seven fold integral

$$\langle \hat{H}_0 \, \hat{a}^\dagger(z)\hat{a}(z)\,\hat{V}\rangle \; = \; 6 \int \ldots \int dz_1 \ldots dz_7 \; V_1(z_1,z_2) V_2(z_3,z_4,z_5,z_6,z_7) \times$$
$$\times \; \langle \hat{a}^\dagger(z_1)\hat{a}(z_2)\hat{a}^\dagger(z)\hat{a}(z)\hat{a}^\dagger(z_4)\hat{a}^\dagger(z_5)\hat{a}(z_6)\hat{a}(z_7)\rangle \qquad (A4)$$

If we then insert into the initial state of Eq. (A1) the definition of the spatial eigenstate $|z;\hat{a}\rangle$ from Eq. (2.3), we obtain a four-fold integral

$$|\psi(t=0)\rangle \; =$$
$$lim_{\Delta 1 \to 0} \, lim_{\Delta 2 \to 0} \iint dz_8 \ldots dz_{11} \; G(z_8-x) \; G(z_9-y) \; s_{\Delta 1}(z_{10}-z_8) \; s_{\Delta 2}(z_{11}-z_9) \; \hat{a}^\dagger(z_{10})\hat{a}^\dagger(z_{11}) \; |vac\rangle (A5)$$

If we insert this initial state into both sides of the expectation value in Eq. (A4), we obtain a 15-fold integral containing the vacuum expectation value of 12 operators.

$$\langle vac| \; \hat{a}(z_{13})\hat{a}(z_{12})\hat{a}^\dagger(z_1)\hat{a}(z_2)\hat{a}^\dagger(z)\hat{a}(z)\hat{a}^\dagger(z_4)\hat{a}^\dagger(z_5)\hat{a}(z_6) \; \hat{a}(z_7)\hat{a}^\dagger(z_{10})\hat{a}^\dagger(z_{11}) \; |vac\rangle \qquad (A6)$$

After making a very frequent and systematic use of the commutator relationship $[\hat{a}(z_1), \, \hat{a}^\dagger(z_2)]_- = \delta\Box z_1-z_2\Box\Box\Box\Box\Box$e 15-fold integral can be reduced to the following cumbersome final form:

$$\langle \hat{H}_0 \, \hat{a}^\dagger(z)\hat{a}(z)\,\hat{V}\rangle \; =$$

$+ 8 \; G(x-z)\int d\xi \; L_{1/2}(\xi-z) \int dz_1 \; L_{1/2}(\xi-z_1) \int dz_2 \; V_1(z_2,z_1)G(y-z_2) \int dz_3 \; L_{1/2}(\xi -z_3)G(x-z_3) \int dz_4 \; L_{1/2}(\xi -z_4)G(y-z_4)$

$+ 8 \; G(y-z)\int d\xi \; L_{1/2}(\xi-z) \int dz_1 \; L_{1/2}(\xi-z_1) \int dz_2 \; V_1(z_2,z_1)G(x-z_2) \int dz_3 \; L_{1/2}(\xi -z_3)G(x-z_3) \int dz_4 \; L_{1/2}(\xi$

-z_4)G(y-z_4)

+ 8 \int d\xi\ L_{1/2}(\xi-z) \int dz_2\ V_1(z_2,z)G(y-z_2) \int dz_4\ L_{1/2}(\xi -z_4)G(y-z_4) \left( \int dz_1\ L_{1/2}(\xi-z_1)G(x-z_1) \right)^2

+ 8 \int d\xi\ L_{1/2}(\xi-z) \int dz_2\ V_1(z_2,z)G(x-z_2) \int dz_4\ L_{1/2}(\xi -z_4)G(x-z_4) \left( \int dz_1\ L_{1/2}(\xi-z_1)G(y-z_1) \right)^2 \qquad (A7)

The derivation of the second term with $\langle (\hat{V} \hat{H}_0 + \hat{H}_0 \hat{V}) \hat{a}^\dagger(z) \hat{a}(z) \rangle$ is similarly cumbersome, and we only state here the final expression

$\langle (\hat{V} \hat{H}_0 + \hat{H}_0 \hat{V})\ \hat{a}^\dagger(z) \hat{a}(z) \rangle$ + c.c. $= 48 \iiiint dz_1 dz_2\ dz_3\ d\xi_1\ d\xi_2\ \{$

$G(x-z_1)\ G(y-\xi_1)\ V_1(z_1,z_2)\ L_{1/2}(z_3-\xi_1)\ L_{1/2}(z_3-z_2)\ L_{1/2}(z_3-z)\ L_{1/2}(z_3-\xi_2)\ G(x-z)\ G(y-z_2)$

$+\ G(x-z_1)\ G(y-\xi_1)\ V_1(z_1,z_2)\ L_{1/2}(z_3-\xi_1)\ L_{1/2}(z_3-z_2)\ L_{1/2}(z_3-z)\ L_{1/2}(z_3-\xi_2)\ G(y-z)\ G(x-\xi_2)$

$+\ G(x-\xi_1)\ G(y-z_1)\ V_1(z_1,z_2)\ L_{1/2}(z_3-\xi_1)\ L_{1/2}(z_3-z_2)\ L_{1/2}(z_3-z)\ L_{1/2}(z_3-\xi_2)\ G(x-z)\ G(y-\xi_2)$

$+\ G(x-\xi_1)\ G(y-z_1)\ V_1(z_1,z_2)\ L_{1/2}(z_3-\xi_1)\ L_{1/2}(z_3-z_2)\ L_{1/2}(z_3-z)\ L_{1/2}(z_3-\xi_2)\ G(y-z)\ G(x-\xi_2)$

$+\ G(x-\xi_1)\ G(y-\xi_2)\ V_1(z_1,z_2)\ L_{1/2}(z_3-\xi_2)\ L_{1/2}(z_3-\xi_1)\ L_{1/2}(z_3-z_1)\ L_{1/2}(z_3-z)\ G(x-z)\ G(y-z_2)$

$+\ G(x-\xi_1)\ G(y-\xi_2)\ V_1(z_1,z_2)\ L_{1/2}(z_3-\xi_2)\ L_{1/2}(z_3-\xi_1)\ L_{1/2}(z_3-z_1)\ L_{1/2}(z_3-z_2)\ G(x-z)\ G(y-z_2)$

$+\ G(x-\xi_1)\ G(y-\xi_2)\ V_1(z_1,z_2)\ L_{1/2}(z_3-\xi_2)\ L_{1/2}(z_3-\xi_1)\ L_{1/2}(z_3-z_1)\ L_{1/2}(z_3-z)\ G(y-z)\ G(x-z_2)$

$+\ G(x-\xi_1)\ G(y-\xi_2)\ V_1(z_1,z_2)\ L_{1/2}(z_3-\xi_2)\ L_{1/2}(z_3-\xi_1)\ L_{1/2}(z_3-z_1)\ L_{1/2}(z_3-z_2)\ G(y-z)\ G(x-z_2)\ \}$ \qquad (A8)

## Appendix B

Let us review here how any quantum field theoretical operator $\hat{A}$ is transformed when the corresponding observable $\langle \Psi | \hat{A} | \Psi \rangle$ is measured as $\langle \Psi | \hat{A}(\theta) | \Psi \rangle$ in a different coordinate system that moves with velocity v relative to the original reference frame. For simplicity we introduce here the rapidity parameter $\theta = \tanh(v/c)$ and the usual boost parameter $\gamma \equiv [1-(v/c)^2]^{-1/2} = \text{Cosh } \theta$. The most fundamental transformation law [28,29] is given by the Heisenberg relationship $i\partial\hat{A}/\partial(c\theta) = [\hat{K}, \hat{A}]$, having the formal solution

$$\hat{A}(\theta) = \exp[-i\hat{K}c\theta]\ \hat{A}\ \exp[i\hat{K}c\theta] \qquad (B1)$$

To shorten our notation we rename from now on the propagator for the boost $\hat{B} \equiv \exp[i\hat{K}c\theta]$ and also omit the argument $\theta$ from any operator associated with the lab frame ($\theta=0$). If we are in the Schrödinger picture, a system characterized in the laboratory frame by the Hilbert state $|\Psi\rangle$ [with $\langle\Psi|\Psi\rangle=1$] would be described in a moving frame as $|\Psi;\theta\rangle \equiv \hat{B}|\Psi\rangle$ to guarantee that $\langle\Psi|\hat{A}(\theta)|\Psi\rangle = \langle\Psi;\theta|\hat{A}|\Psi;\theta\rangle$. The boost operator $\hat{K}$ has to satisfy the Poincare relationships $[\hat{K},\hat{H}]=-i\hat{P}$ and $[\hat{K},\hat{P}]=-i\hat{H}/c^2$. As a result, one possible form would be $\hat{K}=-(\hat{Z}\hat{H}+\hat{H}\hat{Z})/(2c^2)$, where $\hat{Z}$ is the center of mass operator. Equivalently, we can therefore also express the position operator as $\hat{Z}=-c^2(\hat{H}^{-1}\hat{K}+\hat{K}\hat{H}^{-1})/2$.

As a side issue, we also note the same Poincare relationships hold for quantum mechanical operators of single-particle wave functions, where $h=[m^2c^4+c^2p^2]^{1/2}$, $p=-i\partial/\partial z$ and the boost generator $k=-\{z[m^2c^4+c^2p^2]^{1/2}+[m^2c^4+c^2p^2]^{1/2}z\}/(2c^2)$. Here $k$ has a very illustrative non-relativistic limit, $k\rightarrow-m$, such that the corresponding boost propagator $\exp[ikc\theta]$ simplifies to $\exp[-imvz]$, which shifts the momentum of a state by $-mv$, i.e. $\exp[ikc\theta]\,|p\rangle = |p-mv\rangle$.

It turns out that the formal solution Eq. (B1) for some specific set of operators $\hat{A}(\theta)$ can be simplified to explicit expressions in terms of the original operators seen from the original reference frame. These operators are the total momentum, total energy and center of mass operators $\hat{P}=\int dp\ p\ \hat{a}(p)^\dagger\hat{a}(p)$, $\hat{H}=\int dp\ \omega(p)\ \hat{a}(p)^\dagger\hat{a}(p)$ and $\hat{Z}=\int dz\ z\ \hat{a}(z)^\dagger\hat{a}(z)/\int dz\ \hat{a}(z)^\dagger\hat{a}(z)$. The denominator of the latter operator is required to guarantee that $[\hat{Z},\hat{P}]=i$. Had we omitted it we would have obtained the position operator for the 1-particle sector of the Fock space leading to the commutator $[\int dz\ z\ \hat{a}(z)^\dagger\hat{a}(z),\ \hat{P}] = i\int dz\ \hat{a}(z)^\dagger\hat{a}(z)$. The boost–transformed operators are

$$\hat{P}(\theta) = \hat{P}\ \text{Cosh}\theta - \hat{H}/c\ \text{Sinh}\theta \qquad\qquad (B2a)$$

$$\hat{H}(\theta) = \hat{H}\ \text{Cosh}\theta - \hat{P}\ c\ \text{Sinh}\theta \qquad\qquad (B2b)$$

$$\hat{Z}(\theta) = \tfrac{1}{4}\ [\hat{H}(\theta)^{-1}\hat{Z}\hat{H} + \hat{H}(\theta)^{-1}\hat{H}\hat{Z} + \hat{Z}\hat{H}\hat{H}(\theta)^{-1} + \hat{H}\hat{Z}\hat{H}(\theta)^{-1}] \qquad (B2c)$$

$$\hat{K}(\theta) = \hat{K} \qquad\qquad (B2d)$$

The validity of these solutions can be shown by inserting them into the original Heisenberg equation $i\partial\hat{A}/\partial(c\theta)=[\hat{K},\hat{A}]$. Note that while $\hat{P}(\theta)$ is a function of $\hat{P}$ only, the operator $\hat{Z}(\theta)$ depends on $\hat{Z}$ as well as $\hat{P}$.

Also the expressions for the momentum annihilation operator $\hat{a}(p)$ and the single-particle states with given momentum p can be simplified,

$$\hat{B}\,|p\rangle = [dp(\theta)/dp]^{1/2}\,|p(\theta)\rangle \tag{B3a}$$

$$\hat{B}^{\dagger}\,\hat{a}(p)\,\hat{B} = [dp(-\theta)/dp]^{1/2}\,\hat{a}(p(-\theta)) \tag{B3b}$$

where the function $p(\theta) \equiv p\,\mathrm{Cosh}\theta - [m^2c^4 + p^2c^2]^{1/2}/c\;\mathrm{Sinh}\theta$ is also the solution to the transformation law for the classical momentum, given by the Poisson bracket $dp/dc\theta = \{k,p\}_{Z,p}$ with $k = -zh/c^2$.

The proof for Eq. (B3a) found in most textboks uses first the property of Eq. (B2a) for $\hat{P}(\theta)$ and then the requirement that the unit operator should be invariant. If we start with $\hat{P}|p\rangle = p|p\rangle$ and $\hat{H}|p\rangle = h|p\rangle$, multiply each side with the corresponding functions $\mathrm{Cosh}\theta$ and $\mathrm{Sinh}\theta$ and add up the two equations, using Eq. B(2a) we find immediately that $\hat{P}\,\hat{B}\,|p\rangle = p(\theta)\hat{B}\,|p\rangle$. In other words, any state that is proportional to $\hat{B}\,|p\rangle$ is also an eigenstate of $\hat{P}$ with eigenvalue $p(\theta)$. To complete the proof, we have to find the normalization factor $N_{\theta}(p)$ so that $\hat{B}\,|p\rangle = N_{\theta}(p)\,|p(\theta)\rangle$.

In order to find this factor $N_{\theta}(p)$, we require the in the single-particle space the spectral decomposition of unit operator to be invariant, $1 = \int dp|p\rangle\langle p| = \int dp(\theta)|p(\theta)\rangle\langle p(\theta)|$. The unit operator has to be unchanged under the boost, $1 = \int dp\,\hat{B}\,|p\rangle\langle p|\hat{B}^{\dagger}$. If we substitute the variables from p to $p(\theta)$ and introduce the appropriate Jacobian, we obtain $\int dp(\theta)\,|dp/dp(\theta)|\,\hat{B}\,|p\rangle\langle p|\hat{B}^{\dagger}$. If we define the states $|p(\theta)\rangle$ as $|dp/dp(\theta)|^{1/2}\,\hat{B}\,|p\rangle$, such that the Jacobian is absorbed into the state, the unit-operator takes the (required) invariant form $1 = \int dp(\theta)\,|p(\theta)\rangle\langle p(\theta)|$ and we have derived that $N_{\theta}(p) = |dp(\theta)/dp|^{1/2}$.

The proof of Eq. (B3b) follows similarly based on $|p(\theta)\rangle = \hat{a}(p(\theta))^{\dagger}|vac\rangle$, which models how a moving observer would describe a state that has momentum p in the lab frame. If we replace $|p(\theta)\rangle$ by $|dp/dp(\theta)|^{1/2}\,\hat{B}\,\hat{a}(p)^{\dagger}|vac\rangle$ we can insert $\hat{B}^{\dagger}\,\hat{B}$ before the state $|vac\rangle$. If we use the fact that

the vacuum state should look identical to all observers, $\hat{B} |vac\rangle = |vac\rangle$, we find immediately that $\hat{B}$ $\hat{a}(p) \hat{B}^{\dagger} = [dp(\theta)/dp]^{1/2} \hat{a}(p(\theta))$. If we now switch the sign of the rapidity $\theta$, we obtain Eq. (B3b).

**Appendix C**

As is well known, the Schrödinger field operator $\hat{\varphi}(z)$ has a rather unique simplifying property under the *combined* boost and time-shift transformation. In most textbooks this (Lorentz) transformation property is assumed to be valid from the very beginning, but for our discussion it is important to show how the Lorentz transformation actually follows from the Heisenberg equation Eq. (B1) together with the Poincare relationships.

First we note that in contrast to all previous solutions of Appendix B (where the properties under a time-shift were irrelevant) the transformed field-operator $\hat{\varphi}(z)$ has the unique property. It turns out that the z-dependence of $\hat{\varphi}(z;\theta)$ is directly related to the functional form of the *time*-evolved operator $\hat{\varphi}(z,t)$ evaluated at a specific arguments of t and z.

$$\hat{\varphi}(z;\theta) \equiv \hat{B}^{\dagger} \hat{\varphi}(z) \hat{B} = \hat{\varphi}(z \, Cosh\theta, z/c \, Sinh\theta) \qquad (C1)$$

In other words, the operator $\hat{\varphi}(z)$ when seen from an observing frame can be easily computed by simply replacing its original spatial coordinate by $z \, Cosh\theta$ and then replacing the time t [if we know the temporal dependence of $\hat{\varphi}(z,t)$] by $z/c \, Sinh\theta$. This can be seen if we use Eq. (B3b) for $\hat{a}(p)$ in the momentum expansion for $\hat{\varphi}(z)$ in Eq. (2.2a). If we then switch the integration variable from p to $q\equiv p(-\theta)=pCosh\theta + [m^2c^4+p^2c^2]^{1/2}/c \, Sinh\theta$, the argument in the exponent ipz changes to i(q $Cosh\theta-\omega(q)/c \, Sinh\theta)z$. This expression is identical to i[$-\omega(q)t_{-\theta} + p(q)z_{-\theta}$] if we choose $z_{-\theta} \equiv z$ $Cosh\theta$ and $t_{-\theta} \equiv z/c \, Sinh\theta$. Due to the variable substitution $dp = [dp/dq] \, dq = \omega(p)/\omega(q) \, dq$, the pre-factor $[\omega(q)/\omega(p)]^{1/2}$ of $\hat{a}(q)$ and the important factor $\omega(p)^{-1/2}$ in $\hat{\varphi}(z)$, the resulting integral is identical to the original field expansion, except that now only the parameters z and t need to be replaced by $z_{-\theta}$ and $t_{-\theta}$.

The operator $\hat{a}(z)$ does not have the factor $\omega(p)^{-1/2}$ in its momentum expansion and its transformed expression therefore cannot be simplified. Using the Fourier expansion of $\hat{a}(z)$ and the transformation formula Eq. (B3.b) for â(p) we can derive that

$$\hat{a}(z;\theta) \equiv \hat{B}^{\dagger} \, \hat{a}(z) \, \hat{B} = \int dz' \, F_{\theta}(z_{-\theta} - z', t_{-\theta}) \, \hat{a}(z') \tag{C2a}$$

$$F_{\theta}(z_{-\theta} - z', t_{-\theta}) \equiv (2\pi)^{-1} \int dq \, [\omega(p)/\omega(q)]^{1/2} \exp[-i\omega(q)t_{-\theta} + iq(z_{-\theta} - z')] \tag{C2b}$$

Here the factor $\omega(p) = \omega(p(q)) = \omega[q\text{Cosh}\theta + \omega(q)/c \, \text{Sinh}\theta]$ needs to be evaluated as a complicated function of the momentum q.

For completeness and to make contact with the traditional description found in textbooks, we mention as a side note the boost of the Heisenberg operator $\hat{\varphi}(z,t) = \exp[-i\hat{H}t] \, \hat{\varphi}(z) \exp[i\hat{H}t]$, which corresponds to a *combined* boost *and* time-shift transformations of $\hat{\varphi}(z)$ and leads to $\hat{\varphi}(z,t;\theta)$ = $\hat{B}^{\dagger} \exp[-i\hat{H}t] \, \hat{\varphi}(z) \exp[i\hat{H}t] \, \hat{B}$. Following the same variable transformation and the redefinition of the parameters t and z, we would have found the usual simplification $\hat{\varphi}(z,t;\theta) = \hat{\varphi}(z_{-\theta}, t_{-\theta})$ $= \hat{\varphi}(L_{-\theta}(z,t))$, where now the original parameter t is chosen to be non-zero. Obviously, $(a_{\theta}, b_{\theta}) = L_{\theta}(a,b) \equiv (a \, \text{Cosh}\theta - b/c \, \text{Sinh}\theta, b \, \text{Cosh}\theta - a/c \, \text{Sinh}\theta)$ denotes the usual Lorentz formulas for the parameters z and t. Obviously, for the special case of t=0, we recover Eq. (C1).

As a last issue we would like to point out that in the literature it is always assumed from the beginning that the usual Lorentz formulas also describe the combined time and velocity boost for any *interacting* field theory, but a derivation that is solely based on the Poincare relationships is hard to find. Furthermore, this result seems non-trivial as the form of the boost operator is interaction dependent, while the Lorentz transformations are not. We therefore summarize a brief derivation here to show that even the boost transformation for the interacting field operator simplifies to

$$\hat{\Phi}(z,t;\theta) \equiv \hat{B}^{\dagger} \, \hat{\Phi}(z,t) \, \hat{B} = \hat{\Phi}[L_{-\theta}(z,t)] \tag{C3}$$

where $\hat{\Phi}(z,t)$ denotes the time evolution of the field operator $\hat{\varphi}(z)$ under the full interaction and $\hat{B}$ depends now on the interaction. To be as concrete as possible, we use here our $\hat{\varphi}^4$-system as a specific example. Here the (interaction) dependent boost operator takes the form $\hat{K}_{int} = \hat{K}_0 +$ $\lambda \int dz\, z\, \hat{\varphi}(z)^4$. The time-evolution of the interacting field can be written as $\hat{\Phi}(z,t) = \exp[-i\hat{H}t]\, \hat{\Phi}(z)$ $\exp[i\hat{H}t]$ and the coordinate z can be shifted to zero by introducing the shift operator $\hat{\Phi}(z) =$ $\exp[i\hat{P}z]\, \hat{\Phi}(0)\, \exp[-i\hat{P}z]$.

If we insert the unit operator $\hat{B}\,\hat{B}^\dagger$ four times into Eq. (B5) we obtain:

$$\hat{\Phi}(z,t;\theta) = \hat{B}^\dagger \exp[-it\hat{H}]\, \hat{B}\,\hat{B}^\dagger \exp[i\hat{P}z]\, \hat{B}\,\hat{B}^\dagger\, \hat{\Phi}(0)\, \hat{B}\,\hat{B}^\dagger \exp[-i\hat{P}z]\, \hat{B}\,\hat{B}^\dagger \exp[it\hat{H}]\, \hat{B} \quad (C4)$$

The product of the fifteen operators simplifies considerably after four steps. First, the inner most product, $\hat{B}^\dagger \hat{\Phi}(0)\hat{B}$, is actually identical to $\hat{\Phi}(0)$. At the initial time the interacting field $\hat{\Phi}(0)$ agrees with the free-field $\hat{\varphi}(0)$, taking the form from Eq. (2.2a) $\hat{\varphi}(0)=(4\pi)^{-1/2}c \int dk$ $\omega(k)^{-1/2}[\hat{a}(k)+\hat{a}^\dagger(k)]$. Using the transformation properties of the $\hat{a}(k)$ as shown above, the free boost leaves this field invariant, $\exp[-i\hat{K}_0 c\theta]\hat{\varphi}(0) \exp[i\hat{K}_0 c\theta]=\hat{\varphi}(0)$. Furthermore, as the interacting part of the boost-generator $\lambda \int dz\, z\, \hat{\varphi}(z)^4$ also commutes with $\hat{\varphi}(0)$, we have $\hat{B}^\dagger \hat{\varphi}(0)\hat{B} = \hat{\varphi}(0)$.

The second step involves the product $\hat{B}^\dagger \exp[i\hat{P}z]\, \hat{B}$, which can be simplified to $\exp[i(\hat{P}\mathrm{Cosh}\theta - \hat{H}/c\,\mathrm{Sinh}\theta)z]$ using the general solution Eq. (B2a). The third step is quite similar; here the product simplifies to $\hat{B}^\dagger \exp[-i\hat{H}t]\, \hat{B} = \exp[-i(\hat{H}\mathrm{Cosh}\theta - \hat{P}\, c\,\mathrm{Sinh}\theta)t]$, using Eq. (B2b). As the fourth step, we have to combine these two operators leading to $\exp[i\hat{P}(z\mathrm{Cosh}\theta +ct\,\mathrm{Sinh}\theta)] \exp[-i\hat{H}(t\mathrm{Cosh}\theta +z/c\,\mathrm{Sinh}\theta)]$, which we abbreviate as $\exp[-i\hat{H}t_{-\theta}]$ $\exp[i\hat{P}z_{-\theta}]$. The later step is possible as $\hat{H}$ and $\hat{P}$ commute and z and t are only parameters. After these steps, Eq. (B6) simplifies to

$$\hat{\Phi}(z,t;\theta) = \exp[-i\hat{H}t_{-\theta}] \exp[i\hat{P}z_{-\theta}]\, \hat{\Phi}(0)\, \exp[-i\hat{P}z_{-\theta}] \exp[i\hat{H}t_{-\theta}]$$

$$= \hat{\Phi}(z_{-\theta},t_{-\theta}) \equiv \hat{\Phi}(L_{-\theta}(z,t)) \qquad (C5)$$